\documentclass[journal]{IEEEtran}

\usepackage{comment}
\usepackage[T1]{fontenc}
\usepackage{algorithmic}
\usepackage{algorithm}
\usepackage{color}
\usepackage{graphicx}
\usepackage{mathrsfs}
\usepackage{amsfonts,amssymb}
\usepackage{subcaption}
\usepackage{amsmath}
\usepackage{epstopdf}
\usepackage[justification=centering]{caption}
\ifCLASSINFOpdf
\else
\fi

\usepackage{multirow}
\usepackage{amsmath}
\usepackage{bm}
\usepackage[cmintegrals]{newtxmath}
\begin{document}
	\title{Multi-Agent Deep Reinforcement Learning Based Trajectory Planning for Multi-UAV Assisted Mobile Edge Computing}
	
	\author{Liang Wang,
		Kezhi Wang,
		Cunhua Pan,
		Wei Xu,
		Nauman Aslam and Lajos
		Hanzo, \textsl{Fellow, IEEE}
		\thanks{Corresponding Author: Kezhi Wang}
		\thanks{The work of W. Xu was supported in part by the NSFC under grants 62022026 and 61871109. L. Hanzo would like to acknowledge the financial support of the Engineering and Physical Sciences Research Council projects EP/N004558/1, EP/P034284/1, EP/P034284/1, EP/P003990/1 (COALESCE), of the Royal Society's Global Challenges Research Fund Grant as well as of the European Research Council's Advanced Fellow Grant QuantCom.

			Liang Wang, Kezhi Wang and Nauman Aslam are with the Department
			of Computer and Information Science, Northumbria University, Newcastle upon Tyne, NE1 8ST, U.K., emails: \{liang.wang, kezhi.wang, nauman.aslam\}@northumbria.ac.uk.
			
			Cunhua Pan is with School
			of Electronic Engineering and Computer Science, Queen Mary University
			of London, E1 4NS, U.K., email: c.pan@qmul.ac.uk.
			
			Wei Xu is with the National Mobile Communications Research Lab, Southeast University, Nanjing 210096, China, and also with Purple Mountain Laboratories, Nanjing 211111, China, email: wxu@seu.edu.cn.
			
			Lajos Hanzo is with the School of Electronics and Computer Science, University of Southampton, Southampton, SO17 1BJ, U.K., email: lh@ecs.soton.ac.uk.
		}	
	}
	
\maketitle


\begin{abstract}
An unmanned aerial vehicle (UAV)-aided mobile edge computing (MEC) framework is proposed, where several UAVs having different trajectories fly over the target area and support the user equipments (UEs) on the ground. We aim to jointly optimize the geographical fairness among all the UEs, the fairness of each UAV' UE-load and the overall energy consumption of UEs. The above optimization problem includes both integer and continues variables and it is challenging to solve. To address the above problem, a multi-agent deep reinforcement learning based trajectory control algorithm is proposed for managing the trajectory of each UAV independently, where the popular Multi-Agent Deep Deterministic Policy Gradient (MADDPG) method is applied. 
Given the UAVs' trajectories, a low-complexity approach is introduced for optimizing the offloading decisions of UEs. We show that our proposed solution has considerable performance over other traditional algorithms, both in terms of the fairness for serving UEs, fairness of UE-load at each UAV and energy consumption for all the UEs.
\end{abstract}

	
\IEEEpeerreviewmaketitle

\section{Introduction}

\IEEEPARstart{A}{s} a benefit of their compelling features, unmanned aerial vehicles (UAVs) are expected to play a vital role in wireless communication systems. To elaborate a little further, UAVs are capable of providing wireless connectivity even without network infrastructure, or complement the conventional base stations (BSs), whose coverage may suffer from severe blockage due to tall buildings or by the damage caused by natural disasters~\cite{zeng2016wireless}. In order to support reliable communication links, UAVs can promptly adjust their locations according to the dynamic communication environment. Furthermore, since UAVs can be deployed freely and flexibly in three-dimensional (3D) space, direct line-of-sight (LoS) communication with ground-UEs can be readily established, which can potentially boost the throughput in practical scenarios~\cite{8438896}. As a benefit of the above appealing features, in~\cite{7959158} and \cite{fan2018optimal}, both fixed-wing UAVs and rotary-wing UAVs were considered as the relaying nodes, for providing seamless connectivity. In~\cite{8708930}, Wang~\emph{et al.} investigated a fixed-wing UAV-to-UAV communication system, and they proposed a path planning algorithm for minimizing the latency of information transmission, under the constraints of accelerations, location uncertainties and throughput. In~\cite{8685130}, Cui~\emph{et al.} studied the problem of maximizing the average data rate among UEs in mobile-UAV-enabled networks both in orthogonal multiple access (OMA) and non-orthogonal multiple access (NOMA) modes. Furthermore, in agricultural applications, as well as in weather monitoring and wildfire management, UAV can be utilized as a mobile data collector~\cite{lyu2018uav}. As a future development, in~\cite{8365881}, the authors deployed the UAV as the mobile energy transmitter (ET) in a wireless power transfer (WPT) system.

In order to fully exploit the potential of UAVs in wireless communication systems, it is important to investigate their path planning, hovering altitude and trajectory control~\cite{9014313, 8727504, 8489991}. In~\cite{8489991}, Wang~\emph{et al.} creatively proposed a joint UAV altitude and power allocation optimization method, which beneficially alleviated the inter-cell interference of each UAV network.
In~\cite{6863654}, Al-Hourani~\emph{et al.} optimized the latitude of UAVs in order to provide the maximum radio coverage area on the ground. In~\cite{7412759}, both static and  mobile UAVs were considered in device-to-device (D2D) networks. Additionally, the UAV's altitude was optimized for maximizing the system's sum-rate and coverage probability. To tackle the throughput maximization problem of UAV-aided mobile relaying systems, Zeng \emph{et al.}~\cite{zeng2016throughput} proposed an iterative algorithm to optimize the UAV's trajectory and power allocation. In the content of multi-UAV enabled multiuser systems, Wu \emph{et al.}~\cite{8247211} maximized the minimum throughput over all ground users by jointly optimizing the user scheduling, power control and UAV trajectories. In order to meet the different quality-of-service (QoS) requirement of users, Alzenad~\emph{et al.}~\cite{7918510} investigated coverage-placement problem of UAV-BSs and proposed an optimal placement algorithm for maximizing the number of users supported.

In recent years, mobile edge computing (MEC) has been shown to dramatically improve the user experience~\cite{hu2015mobile, 8754787}. By providing both computing and storage hardware at the network edges, namely at the BSs or access points (APs), the resource-limited UEs have the option of offloading their computation-intensive and latency-critical applications to the MEC servers~\cite{8016573}. Due to the mobility of UAVs, recent years have seen research progress on the integration of UAVs with MEC~\cite{yang2019energy, 8873672}. In~\cite{motlagh2017uav}, Motlagh~\emph{et al.}, were amongst the first who proposed UAV-enabled MEC, in which UEs can significantly reduce the energy consumption via offloading. In order to minimize the overall energy dissipation of UEs while meeting their QoS requirement, Jeong \emph{et al.}~\cite{7932157} proposed an efficient successive convex approximation-based algorithm for jointly optimizing the bit allocation and UAV's trajectory. Considering a multi-UAV system, Hua~\emph{et al.}~\cite{hua2019energy} investigated the multi-UAV scenario, and they optimized the UAVs' trajectories, transmit power and user scheduling.

Given the recent advances in machine learning \cite{8957702}, 
the combination of deep neural networks (DNNs)~\cite{lecun2015deep} and reinforcement learning (RL)~\cite{sutton1998introduction}, i.e., deep reinforcement learning (DRL) has become a hot research topic. In DRL, an agent is assumed to interact with the environment for learning the optimal policy with the aid of exploration. Compared to traditional RL, DRL facilitates more accurate convergence and approximation by exploiting the power of DNNs for estimating the associated functions in RL~\cite{li2017deep}. The great potential of DRL in solving complex control problems has also been demonstrated in~\cite{mnih2015human, 8906180, van2016deep, lillicrap2015continuous, bu2008comprehensive}. In~\cite{mnih2015human}, Mnih \emph{et al.} introduced the deep Q network (DQN) philosophy, which ignited the field of DRL. For instance, Wang~\emph{et al.}\cite{8906180} systematically investigated the problem of distributed Q-learning aided heterogeneous network association in the content of energy-efficient Internet of things (IoT).
In order to improve the training procedure, DQN relies on a pair of techniques namely, experience replay and target networks. For the sake of tackling the typical over-estimation problem of RL, a double DQN (D-DQN) was proposed by Van Hasselt~\emph{et al.}~\cite{van2016deep}. However, DQN may suffer from the curse of high-dimensional action spaces and cannot be readily applied to continuous domains. Thus, motivated by this, Lillicrap~\emph{et al.}~\cite{lillicrap2015continuous} proposed a deep deterministic policy gradient (DDPG) technique based on the so-called actor-critic architecture, which can be readily applied for a range of challenging problems. A comprehensive survey of multi-agent RL, have also been provided by Bu~\emph{et al}~\cite{bu2008comprehensive}.

Against the above background, we conceive a multi-UAV assisted MEC framework, where each UAV is controlled by a dedicated agent. We aim for jointly maximizing the geographical fairness$\footnote{The geographical fairness reflects the QoS level of UEs served by UAVs from the initial time slot to the current time slot. }$ among the UEs covered, the fairness of UE-load of each UAV$\footnote{The UE-load of UAV is defined in (\ref{cmt}).}$, while minimizing the overall energy consumption of UEs by optimizing each UAV's trajectory and offloading decisions. This is a complex problem which includes 
both integer and continuous variables. Hence it is challenging to address it by traditional algorithms, such as convex optimization and dynamic programming. Therefore, we conceive a multi-agent deep reinforcement learning based solution, with the help of the popular Multi-Agent Deep Deterministic Policy Gradient (MADDPG) \cite{10.5555/3295222.3295385} for solving it. 
Given the UAVs' trajectories, a low-complexity approach is introduced for optimizing the offloading decisions of UEs.
Our simulation results will show that the proposed DRL based algorithm outperforms the benchmark algorithms. We summarize the difference between our work and the existing
literature in Table~\ref{comp}.
\begin{table*}[htbp!]
	\caption{Comparison between our work and the existing
		literature. }
	\begin{center}
		\begin{tabular}{|c|c|c|c|c|c|c|c|c|}
			\hline
			\multirow{1}{*}{Reference} & \shortstack{Single \\ UAV}&\shortstack{Multi \\ UAV}&\shortstack{Mobile edge \\ computing (incl.)}&\shortstack{Path \\ planning}&\shortstack{Offloading \\ decision}&\shortstack{Reinforcement learning \\ (e.g., Q-learning)}&\shortstack{Multi \\ agent learning}&\shortstack{DNN}\\
			\hline
			
			\cite{8438896} &\checkmark&&&\checkmark&&&& \\  \hline
			
			\cite{fan2018optimal} &\checkmark&&&\checkmark&&&& \\  \hline
			
			\cite{8708930} &&\checkmark&&\checkmark&&&& \\  \hline
			
			\cite{8685130} &\checkmark&&&\checkmark&&&& \\  \hline
			\cite{lyu2018uav} &\checkmark&&&\checkmark&&&& \\  \hline
			\cite{8365881}&\checkmark&&&\checkmark&&&& \\  \hline
			\cite{9014313}&&\checkmark&&\checkmark&&&&\checkmark \\  \hline
			\cite{8727504}&&\checkmark&&\checkmark&&\checkmark&\checkmark& \\  \hline
			\cite{8489991}&&\checkmark&&\checkmark&&&& \\ \hline
			\cite{7412759}&\checkmark&&&\checkmark&&&& \\  \hline
			\cite{zeng2016throughput}&&\checkmark&&\checkmark&&&& \\  \hline
			\cite{8247211}&&\checkmark&&\checkmark&&&& \\  \hline
			\cite{7918510}&\checkmark&&&\checkmark&&&& \\  \hline
			\cite{yang2019energy}&&\checkmark&\checkmark&\checkmark&\checkmark&&& \\  \hline
			\cite{8873672}&\checkmark&&&\checkmark&\checkmark&&& \\  \hline
			\cite{7932157}&\checkmark&&\checkmark&\checkmark&&&& \\  \hline
			\cite{hua2019energy}&&\checkmark&&\checkmark&&&& \\  \hline
			\cite{8906180}&&&&&&\checkmark&\checkmark& \\  \hline
			\cite{10.5555/3295222.3295385}&&&&&&\checkmark&\checkmark&\checkmark \\  \hline		
			Our work& &\checkmark&\checkmark&\checkmark&\checkmark&\checkmark&\checkmark&\checkmark \\  \hline
		\end{tabular}
		\label{comp}
	\end{center}
\end{table*}

The rest of the paper is organized as follows. In Section \uppercase\expandafter{\romannumeral2}, we introduce the system model and the optimization problem. In Section \uppercase\expandafter{\romannumeral3}, our multi-agent based DRL algorithm is proposed. Our experimental results are shown in Section \uppercase\expandafter{\romannumeral4}. Finally, our conclusions are drawn in Section \uppercase\expandafter{\romannumeral5}.

The main notations used in this paper are summarized in Table~\ref{tab1}.

\begin{table}[htbp!]
	\caption{List of main notations}
	\begin{center}
		\begin{tabular}{|l|r|}
			\hline
			\textbf{Notation}&\textbf{Description} \\
			\hline
			n, N, $\mathcal{N}$& The index, number and the set of UEs\\
			\hline
			m, M, $\mathcal{M}$& The index, number and the set of UAVs\\
			\hline
			t, T, $\mathcal{T}$& The index, number and the set of TSs\\
			\hline
			$z_{n,m,t}$ & Offloading decision of UE $n$\\
			\hline
			$S_{n,t}$& Computation task of UE $n$ in TS $t$\\
			\hline
			$D_{n,t}$& Data volume of task $S_{n,t}$\\
			\hline
			$F_{n,t}$& Overall CPU cycles required for task $S_{n,t}$\\
			\hline
			$f_{n,m,t}$& Computation capacity of UAV $m$ allocated to UE $n$\\
			\hline
			$T_{n,m,t}^C$& Execution time of UAV $m$ to UE $n$ in TS $t$ \\
			\hline
			$T_{n,m,t}^{Tr}$& Transmission time of UE $n$ to UAV $m$ in TS $t$  \\
			\hline
			$T^{max}$& Maximal time duration of each TS\\
			\hline
			$\alpha_{m,t}, d_{m,t}$& Flying angle and distance  of UAV $m$ in TS $t$ \\
			\hline		
			$d^{max}$& Maximal flying distance of UAV in each TS\\
			\hline
			$[X_{m,t},Y_{m,t},H]$& Coordinates of UAV $m$ in TS $t$\\
			\hline
			$[x_n,y_n]$& Coordinates of UE $n$\\
			\hline
			$R_{n,m,t}$& Horizontal distance between UAV $m$ and UE $n$\\
			\hline
			$R_{m,m',t}$& Horizontal distance between UAV $m$ and UAV $m'$\\
			\hline
			$R^{max}$& Maximal horizontal coverage radius of UAV\\
			\hline
			$r_{n,m,t}$& Transmitting data rate of UE $n$ to UAV $m$\\
			\hline
			$E_{n,m,t}^{C}$& Energy consumption for task execution\\
			\hline
			$E_{n,m,t}^{Tr}$& Energy consumption for offloading\\
			\hline
			$c_{m,t}$& Relative UE-load of UAV $m$ in TS $t$\\
			\hline
			$f^u_t$& Fairness index of UE-load of each UAV in TS $t$\\
			\hline
			$f^e_t$& Fairness index of UEs in TS $t$ \\
			\hline
		\end{tabular}
		\label{tab1}
	\end{center}
\end{table}

\section{System Model}

\begin{figure*}[htpb]
	\centering
	\includegraphics[width=6.5in]{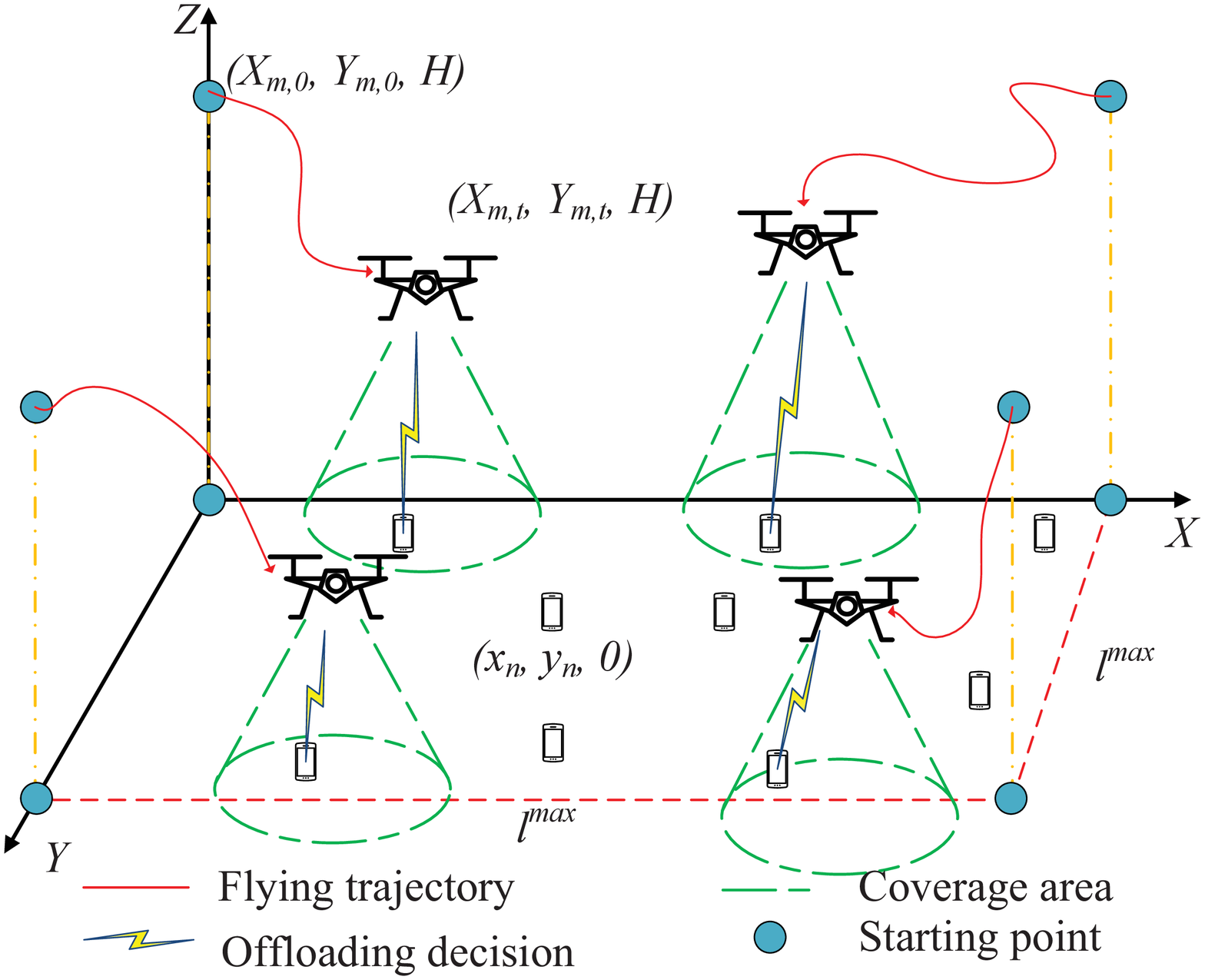}
	\caption{Overall System Architecture}\label{sys}
\end{figure*}

In this section, we describe the system model. As shown in Fig.~\ref{sys}, we assume that there are $N$ UEs randomly distributed in a square-shaped area with side length $l^{max}$, and the set of UEs is denoted as $\mathcal{N} \triangleq \{n=1,2,...N\}$. There are $M$ UAVs flying at a fixed altitude $H$ over the target area to serve the ground UEs, and the set of UAVs is denoted as $\mathcal{M} \triangleq \{m=1,2,...,M \}$. We also assume that UAVs can be deployed and easily charged on the building roof when UAVs run out of their energy. Assume that each UE has a computational task to be executed at each time slot (TS) over $T$ consecutive TSs, $\mathcal{T} \triangleq \{t=1,2,...,T\}$. Each of the tasks can be executed either by the UE or offloaded to one of the UAVs. We define a new set $m \in \mathcal{M}' \triangleq \{0,1,...,M\}$ to denote the possible places where the tasks can be executed, with $m=0$ representing local execution. Then, we define the offloading decision variable $z_{n,m,t}$ as
\begin{equation}\label{c1}
\begin{aligned}
z_{n,m,t} = \{0,1\}, \forall n \in \mathcal{N}, m \in \mathcal{M'}, t \in \mathcal{T},
\end{aligned}
\end{equation}
where $z_{n,m,t} = 1, m \neq 0$ means that UE $n$ decides to offload the task to UAV $m$ in TS $t$, while $z_{n,m,t} = 1, m = 0$ represents that UE $n$ carries out the task itself in TS $t$, and otherwise $z_{n,m,t} = 0$. Furthermore, we assume that each task can only be executed at a single place. Thus, we have
\begin{equation}\label{c2}
\begin{aligned}
\sum_{m=0}^{M} z_{n,m,t} = 1, \forall n \in \mathcal{N}, t \in \mathcal{T}.
\end{aligned}
\end{equation}

Similarly to~\cite{7393804}, in the TS $t$, we assume that UE $n$ has a computationally intensive task $S_{n,t}$ to be executed, which is defined as
\begin{equation}
\begin{aligned}
S_{n,t} = \{D_{n,t}, F_{n,t}\}, \forall n \in \mathcal{N}, t \in \mathcal{T},
\end{aligned}
\end{equation}
where $D_{n,t}$ denotes the data volume to be processed, while $F_{n,t}$ describes the total number of the CPU cycles required for executing this task. Both $D_{n,t}$ and $F_{n,t}$ can be characterized as in~\cite{yang2013framework}.

Furthermore, in TS $t$, each of UAV flies in a direction determined by the angle of $\alpha_{m,t} \in [0, 2\pi)$, distance of $d_{m,t} \in [0, d^{max}]$, and cannot go beyond the border of the target area. We assume that the initial coordinates of UAV $m$ are set as $[X_{m,0}, Y_{m,0}, H]$. Then, the coordinates of UAV $m$ in TS $t$ can be calculated as $[X_{m,t}, Y_{m,t}, H]$, where $X_{m,t}=X_{m,0}+\sum_{t'=1}^{t}d_{m,t'}\text{cos}(\alpha_{m,t'})$ and $Y_{m,t}=Y_{m,0}+\sum_{t'=1}^{t}d_{m,t'}\text{sin}(\alpha_{m,t'})$. Thus, we have
\begin{equation}
\begin{aligned}
	0 \leq X_{m,t} \leq l^{max},~\forall m \in \mathcal{M}, t \in \mathcal{T},
\end{aligned}
\end{equation}
and 
\begin{equation}
	\begin{aligned}
	0 \leq Y_{m,t} \leq l^{max},~\forall m \in \mathcal{M}, t \in \mathcal{T}.
	\end{aligned}
\end{equation}

Additionally, we denote the distance between UAV $m$ and UAV $m'$ in TS $t$ as $R_{m,m',t}$, which can be expressed as
\begin{equation}
\begin{aligned}
R_{m,m',t} = \sqrt{(X_{m,t}-X_{m',t})^2+(Y_{m,t}-Y_{m',t})^2}.
\end{aligned}
\end{equation}
We assume that the UAVs should keep a minimal distance of $R^u$ for avoiding their collision in each TS. Then, we have
\begin{equation}
\begin{aligned}
R_{m,m',t} \geq R^u, \forall m, m' \in \mathcal{M}, m \neq m'.
\end{aligned}
\end{equation}

The horizontal distance between UE $n$ and UAV $m$ in TS $t$ is calculated as
\begin{equation}
\begin{aligned}
R_{n,m,t}=&\sqrt{(X_{m,t}-x_n)^2+(Y_{m,t}-y_n)^2},\\ &\quad \quad~\forall n \in N, m \in \mathcal{M}, t \in \mathcal{T}, \\
\end{aligned}
\end{equation}
where $[x_n, y_n]$ is assumed to be the coordinate of UE $n$. Note that if UE $n$ decides to offload a task to UAV $m$ in TS $t$, it must be in the coverage of UAV $m$. Then, we have
\begin{equation}
\begin{aligned}\label{c4}
z_{n,m,t}R_{n,m,t} \leq R^{max},~\forall n \in \mathcal{N}, m \in \mathcal{M}, t \in \mathcal{T},
\end{aligned}
\end{equation}
where $R^{max}$ is the maximal horizontal coverage radius of the UAVs.

Then, the offloading data rate can be expressed by
\begin{equation}
\begin{aligned}
r_{n,m,t} =& B\text{log}_2\left(1+\frac{\rho P_n}{H^2+R_{n,m,t}^2}\right),\\& \quad \quad~\forall n \in \mathcal{N}, m \in \mathcal{M}, t \in \mathcal{T},
\end{aligned}
\end{equation}
where $B$ is the channel's bandwidth, $P_n$ is the transmission power of UE $n$, $\rho =g_0G_0/\sigma^2$, $G_0 \approx 2.2846$, $g_0$ is the channel's power gain at the reference distance of 1 $m$ and $\sigma^2$ is the noise power~\cite{he2017joint}. Here we do not consider any
particular modulation and coding scheme.

Thus, if UE $n$ decides for offloading its task to UAV $m$ in TS $t$, the time required for offloading the data is given by
\begin{equation}
\begin{aligned}
T^{Tr}_{n,m,t} = \frac{D_{n,t}}{r_{n,m,t}},~\forall n \in \mathcal{N}, m \in \mathcal{M}, t \in \mathcal{T},
\end{aligned}
\end{equation}
and the execution time of the task can be expressed as
\begin{equation}
\begin{aligned}
T_{n,m,t}^{C} = \frac{F_{n,t}}{f_{n,m,t}}, \forall n \in \mathcal{N}, m \in \mathcal{M}', t \in \mathcal{T},
\end{aligned}
\end{equation}
where $f_{n,m,t}$ represents the computational capability of UAV $m$ that can be allocated to UE $n$, and $m=0$ indicates local execution.
Thus, the overall time required for executing the task can be described as
\begin{equation}
\begin{aligned}[l]
T_{n,m,t} = 
\begin{cases}
T_{n,m,t}^{C}, & \text{if local execution}, \\
T_{n,m,t}^{Tr}+T_{n,m,t}^C, & \text{if offloading}. \\
\end{cases}
\end{aligned}
\end{equation}

We also assume that all tasks should be executed within the maximal time duration $T^{max}$ of TS. Then, we have 
\begin{equation}\label{c3}
\begin{aligned}
z_{n,m,t}T_{n,m,t} \leq T^{max},~\forall n \in \mathcal{N}, m \in \mathcal{M}', t \in \mathcal{T}. 
\end{aligned}
\end{equation}

According to~\cite{8274943}, if the UE $n$ decides to execute a task locally, the energy consumption is given by
\begin{equation}
\begin{aligned}
E_{n,m,t}^C=k_n(f_{n,m,t})^{v_n}T_{n,m,t}^C,~\forall n \in \mathcal{N}, t \in \mathcal{T},
\end{aligned}
\end{equation}
where $k_n \geq 0$, $v_n \geq 1$ are positive coefficients.

If UE $n$ decides to offload a task, the energy consumption of offloading is
\begin{equation}
\begin{aligned}
E_{n,m,t}^{Tr} = P_nT_{n,m,t}^{Tr},~\forall n \in \mathcal{N}, m \in \mathcal{M}, t \in \mathcal{T}.
\end{aligned}
\end{equation}
Thus, the energy consumption at UE $n$ can be expressed as
\begin{equation}\label{energy}
\begin{aligned}[l]
E_{n,m,t} = 
\begin{cases}
E_{n,m,t}^C, & \text{if local execution}, \\
E_{n,m,t}^{Tr}, & \text{if offloading}. \\
\end{cases}
\end{aligned}
\end{equation}

Then, we define $c_{m,t} \in [0,1]$ as the relative UE-load of UAV $m$ in TS $t$, as:
\begin{equation}\label{cmt}
\begin{aligned}
c_{m,t} = \frac{\sum_{n=1}^{N}z_{n,m,t}}{N},~\forall m \in \mathcal{M}, t \in \mathcal{T}.
\end{aligned}
\end{equation}
In this paper, our first objective is to minimize the total energy consumption of UEs via optimizing both the offloading decisions and the UAVs' trajectories. However, this may lead to an unfair process since some UAVs may serve more UEs than others. To address this issue, we define a fairness index $f^u_t$ as
\begin{equation}
\begin{aligned}
f^u_t = \frac{\big(\sum_{m=1}^{M}\sum_{t'=1}^{t}c_{m,t'}\big)^2}{M\sum_{m=1}^{M}\big(\sum_{t'=1}^{t}c_{m,t'}\big)^2},
\end{aligned}
\end{equation}
where $f^u_t$ reflects the level of fairness among the UAVs physically, if all the UAVs have a similar UE-load commencing from the initial TS up to TS $t$, the value of $f^u_t$ is closer to 1.

Then, to avoid the situation that some UEs are served during many TSs, while others are never served at all, we define another geographical fairness $f^e_t$ as follows
\begin{equation}
\begin{aligned}
f^e_t = \frac{\big(\sum_{n=1}^{N}\sum_{t'=1}^{t}z_{n,m,t'}\big)^2}{N\sum_{n=1}^{N}\big(\sum_{t'=1}^{t}z_{n,m,t'}\big)^2},
\end{aligned}
\end{equation}
where $f^e_t$ reflects the level of fairness among the UEs, explicitly, if all UEs are served for a similar number of TSs commencing from the initial TS to the TS $t$, the value of $f^e_t$ is closer to 1.

Then, we formulate our optimization problem as follows
\begin{subequations}\label{all}
	\begin{IEEEeqnarray}{s,lCl'lCl'lCl}
		& \IEEEeqnarraymulticol{9}{l}{\mathcal{P}1: \underset{\bm{P},\bm{Z}}{\text{max}}~ 
			\sum_{t=1}^{T}\frac{f^u_t \cdot f^e_t}{\sum_{n=1}^{N}\sum_{m=0}^{M}z_{n,m,t}E_{n,m,t}}} \\
		& \text{subject to:} \nonumber\\
		& z_{n,m,t} = \{0,1\}, \forall n \in \mathcal{N}, m \in \mathcal{M'}, t \in \mathcal{T}, \\
		& \sum_{m=0}^{M} z_{n,m,t} = 1, \forall n \in \mathcal{N}, t \in \mathcal{T}, \\
		& 0 \leq X_{m,t} \leq l^{max},~\forall m \in \mathcal{M}, t \in \mathcal{T}, \\
		& 0 \leq Y_{m,t} \leq l^{max},~\forall m \in \mathcal{M}, t \in \mathcal{T}, \\
		& 0 \leq \alpha_{m,t} <  2\pi,~\forall m \in \mathcal{M}, t \in \mathcal{T}, \\
		& 0 \leq d_{m,t} \leq d^{max},~\forall m \in \mathcal{M}, t \in \mathcal{T},\\
		& R_{m,m',t} \geq R^u, \forall m, m' \in \mathcal{M}, m \neq m', \\
		& z_{n,m,t}R_{n,m,t} \leq R^{max},~\forall n \in \mathcal{N}, m \in \mathcal{M}, t \in \mathcal{T}, \\
		& z_{n,m,t}T_{n,m,t} \leq T^{max},~\forall n \in \mathcal{N}, m \in \mathcal{M}', t \in \mathcal{T}.
	\end{IEEEeqnarray}
\end{subequations}
where $\bm{P} = \{\alpha_{m,t}, d_{m,t}, \forall m \in \mathcal{M}, t \in \mathcal{T}\}$ and $\bm{Z} = \{z_{n,m,t}, \forall n \in \mathcal{N}, m \in \mathcal{M'}, t \in \mathcal{T}\}$. Our objectives are to maximize the fairness of UE-load of each UAV and the fairness of the number of times that each UE is served by UAVs over all the TSs, while minimizing the overall energy consumption of UEs. It is readily observed that the optimization problem cannot be solved by traditional approaches, since it involves both the continuous variables $\bm{P}$ and the discrete variables $\bm{Z}$. Thus, in this paper, a Multi-Agent deep reinforcement learning based Trajectory control algorithm (MAT) is proposed.

\section{The Proposed Algorithm}
In this section, we present our proposed algorithm. First, some background knowledge on deep reinforcement learning is provided, followed by our MAT conceived for solving Problem (\ref{all}).
\subsection{Background Knowledge}
In the traditional reinforcement learning setup, a Markov decision process (MDP)~\cite{bertsekas1995dynamic} is employed with the state space of $\mathcal{S} = \{s_t= s_1,s_2,...,s_T\}$ and the action space of $\mathcal{A} = \{a_t=a_1,a_2,...,a_T\}$. In the MDP, a decision agent interacts with the environment in discrete TSs. More specifically, the decision agent observes the current state $s_t$ of the environment and takes the action $a_t$ that is allowed in that state. As a benefit, the agent will obtain a reward $r_t$ and traverses to a new state $s_{t+1}$. In~\cite{lillicrap2015continuous}, a policy $a_t= \pi(s_t)$ is introduced that maps the state to a legitimate action. During the process of interacting with the environment, the agent aims for selecting the beneficial policy that maximizes the accumulated reward $R_t = \sum_{i=t}^{T}\gamma^{i-t}r_i$, where $\gamma \in (0,1)$ is the discount factor and $T$ is the number of TSs. Additionally, as a beneficial combination of DNN and RL, the philosophy of DQN~\cite{mnih2015human} was proposed, which uses an action-value function $Q(\cdot)$ for approximately evaluating $R_t$ by applying $a_t$, $s_t$ and following $\pi(\cdot)$:
\begin{equation}
\begin{aligned}
Q(s_t,a_t) = \mathbb{E}_{\pi}\big[R_t|s_t,a_t\big],
\end{aligned}
\end{equation}
which is known as the Q-function, and can be obtained by a DNN. Then, the DNN can be trained with the aid of the loss function $L(\cdot)$ defined as:
\begin{equation}\label{lossf}
\begin{aligned}
L(\theta^Q) = \mathbb{E}\big[y_t-Q(s_t,a_t|\theta^Q)^2\big],
\end{aligned}
\end{equation}
where $\theta^Q$ denotes the network parameter of the DNN and $y_t$ is formulated by
\begin{equation}
\begin{aligned}
y_t = r_t + \gamma Q(s_{t+1},a_{t+1}|\theta^Q).
\end{aligned}
\end{equation}
where $a_{t+1}$ denotes the action of the agent in the next TS, which is generated by the DNN, given the state $s_{t+1}$.

Furthermore, in order to make the network more stable, a pair of techniques, namely  the experience replay and the target network are utilized. The experience replay employs a buffer for storing transitions for the sake of mitigating the correlations between consecutive transitions and hence increasing their independence. Compared to the original RL training procedure, the DQN relies on a mini-batch for randomly sampling the transitions from the experience replay buffer, rather than only selecting a single transition. Furthermore, the target network that has the same network structure as $Q(s_t, a_t)$ is employed for reducing the correlations. Note that the target network is only updated at certain intervals.

However, it is proved in~\cite{lillicrap2015continuous} that DQN cannot be directly used for solving continuous-valued control problems. Thus, the popular actor-critic method at DDPG of~\cite{lillicrap2015continuous} is resorted to. Specifically, DDPG consists of a DNN turned as actor and a DQN referred to as the critic network. The actor carries our the mapping function $\pi(s_t|\theta^{\pi})$ while the critic performs the function $Q(s_t,a_t|\theta^Q)$. The actor can generate the optimal action $a_t$ based on the state $s_t$ and it can be trained by applying the policy gradient method of~\cite{silver2014deterministic} defined as
\begin{equation}
\begin{aligned}
\nabla_{\theta^{\pi}}J &\approx \mathbb{E}\big[\nabla_{\theta^{\pi}}Q(s,a|\theta^Q)|_{s=s_t,a=\pi(s_t|\theta^{\pi})}\big]\\
& = \mathbb{E}\big[\nabla_aQ(s,a|\theta^Q)|_{s=s_t,a=\pi(s_t)\nabla_{\theta}^{\pi}\pi(s|\theta^{\pi})|_{s=s_t}}\big],
\end{aligned}
\end{equation} 
while the critic network can be updated by using the loss function of (\ref{lossf}).

\subsection{MAT}
In this section, by applying the popular MADDPG \cite{10.5555/3295222.3295385}, we conceive a multi-agent MDP, namely an observable Markov game~\cite{littman1994markov}. It is assumed that there are $M$ agents interacting with the environment characterized by a set of states $\mathcal{S} \triangleq \{s_t, t \in \mathcal{T}\}$ and a set of actions $\mathcal{A} \triangleq \{a_t, t \in \mathcal{T}\}$. The state $s_t$ consists of the private observation $o_{m,t}$ and some other extra information known by each agent. Additionally, each UAV is controlled by its dedicated agent. In each TS, each agent obtains its private observation $o_{m,t}$ and takes its own action $a_{m,t}$ as well as receives a reward $r_{m,t}$. Then, the environment updates the state and traverses to a new state. Note that each agent is equipped with an actor network $a_{m,t} = \pi^m(o_{m,t})$, a critic network $Q^m(s_t,a_t)$, their target networks $a_{m,t+1}=\pi^{m'}(o_{m,t+1})$ and $Q^{m'}(s_{t+1},a_{t+1})$, as well as an experience replay buffer $B_m$.

The proposed algorithm is based on the framework of centralized training combined with decentralized execution. During the training process, each agent sends its own private observation $o_{m,t}$ and action $a_{m,t}$ to the environment, and then the states $s_t$ which consist of the observations of all the agents and actions are sent back to each agent. Here, all the agents can exchange their private information simultaneously with each other, including coordinates. Furthermore, the critic network of each agent is trained with the states and actions that includes all the agents' observations and actions. Then, during the testing process, each agent can execute its action by only receiving its own private observations $o_{m,t}$, which can potentially maximize the accumulated rewards.
	
Thus, we define the observation, action and reward function for each agent in TS $t$ as follows:

\begin{enumerate}
	\item Observation $o_{m,t}$: we first add the coordinates $[X_{m,t}, Y_{m,t}]$ of UAV $m$ in TS $t$ into the observation of agent $m$. For avoiding collisions between each pair of UAVs, we define the set of relative UAV distances $\{R_{m,m',t},~ m' \in \mathcal{M}, m'\neq m \}$ as part of the observation. Additionally, for better exploration, we also add the set of accumulated times of UEs served by UAVs and UE-load of UAVs commencing from the initial TS up to TS $t$, i.e., $\{\sum_{t'=1}^{t}z_{n,m,t'} ,~\forall n \in \mathcal{N}\}, \{\sum_{t'=1}^{t}c_{m,t'},~\forall m \in \mathcal{M}\}$, respectively into the observation set.	
	\item Action $a_{m,t}$: we define the UAV's flying direction and distance as the action $a_{m,t} = \{\alpha_{m,t}, d_{m,t}\}$ of the $m$-th UAV in the $t$-th TS. 
	\item Reward Function $r_{m,t}$: we define the reward function as:
	\begin{equation}\label{rew_fun}
	\begin{aligned}
	r_{m,t} = \frac{f^u_t \cdot f^e_t}{\frac{1}{N}\sum_{n=1}^{N}\sum_{m=0}^{M}z_{n,m,t} \cdot E_{n,m,t}}- p_m,
	\end{aligned}
	\end{equation}
	where $p_m$ is the penalty incurred if UAV $m$ flies out of the target area or UAV $m$ is collided with another UAV (i.e., the relative distance is under the defined limit).
\end{enumerate}
Then, we define the entire state $s_t$, and action $a_t$ as follows
\begin{enumerate}
	\item State $s_t$: the state consists of the observations of all the agents, which is expressed as $s_t = \{o_{m,t},~\forall m \in \mathcal{M}\}$.
	\item Action $a_t$: the action consists of the actions of all the agents, which is $a_t=\{a_{m,t},~\forall m\in \mathcal{M}\}$.
\end{enumerate}

\begin{figure*}[htpb]
	\centering
	\includegraphics[width=6.5in]{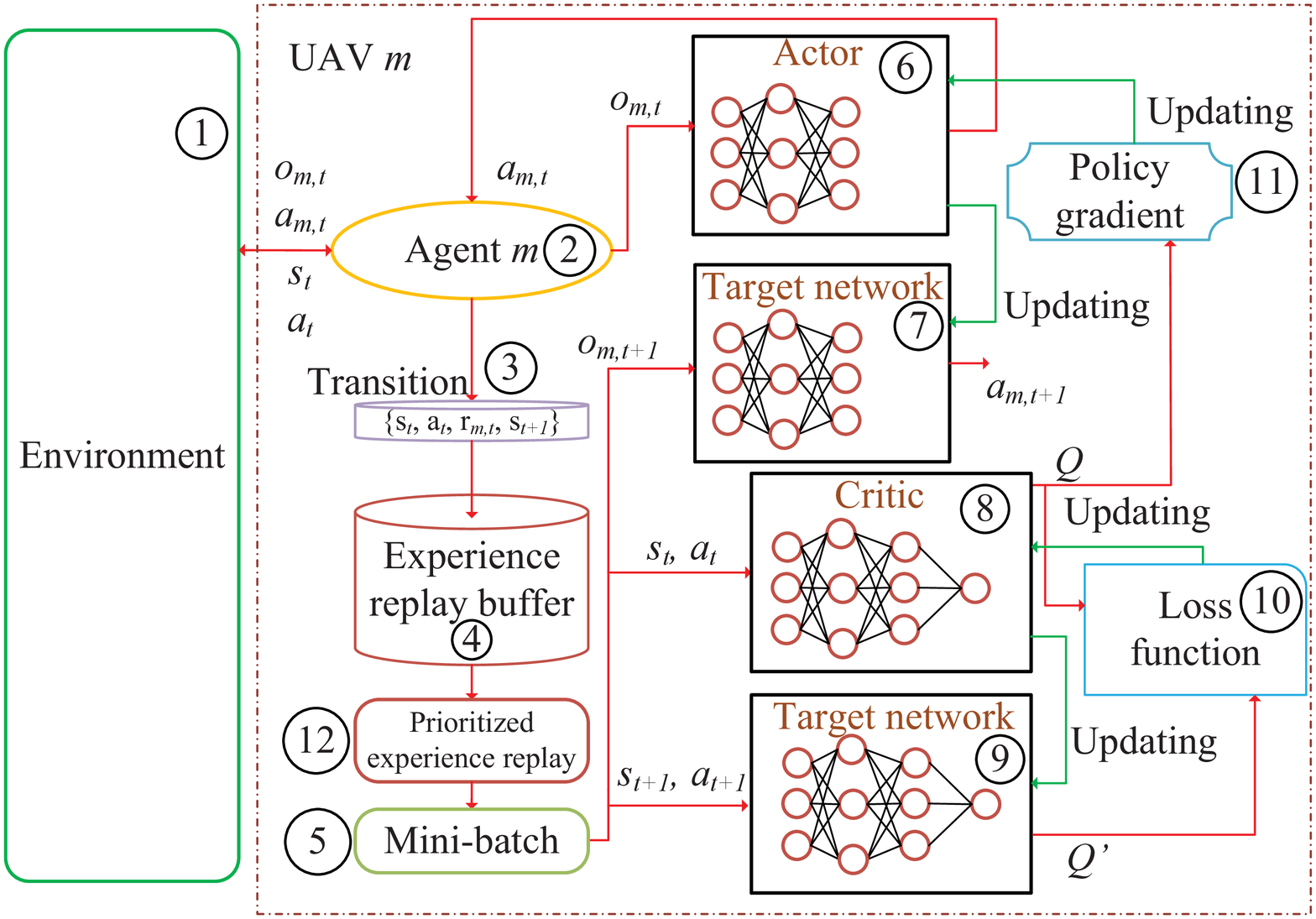}
	\caption{Structure of UAV $m$ (i.e., controlled by Agent $m$)}\label{flow}
\end{figure*}

We show the structure of agent $m$ in Fig.~\ref{flow}. During its interaction with the environment \textcircled{\small{1}}, each UAV (controlled by agent) \textcircled{\small{2}} selects the optimal action associated with its actor network $\pi^m(\cdot)$ \textcircled{\small{6}}, and then obtains the Q value from the critic network $Q^m(\cdot)$ \textcircled{\small{8}} as well as its target action and target Q value from $\pi^{m'}$ \textcircled{\small{7}} and $Q^{m'}(\cdot)$ \textcircled{\small{9}} respectively. The profile of observation, action and reward, which determine the transition are defined as $e_{m,t} \triangleq \{s_t,a_t,r_{m,t},s_{t+1}\}$ that are stored in the experience replay buffer \textcircled{\small{4}}. However, during the training procedure, randomly sampling the mini-batch \textcircled{\small{5}} may have unpredictable effects, since some transitions associated with poor attempts may lead to the termination of the training procedure or may not converge. As a result, Schaul~\emph{et al}.~\cite{schaul2015prioritized} pointed out that transitions having high Temporal Difference (TD)-error often indicate successful attempts. The TD-error $\delta_{m}$ of agent $m$ can be defined as follows
\begin{equation}
\begin{aligned}
\delta_{m} = & r_{m,t}+\gamma Q^{m'}(s_{t+1},a_{t+1}|\theta^{Q^{m'}}) \\& - Q^{m}(s_t,a_{t}|\theta^{Q^m}),~\forall m \in \mathcal{M}, t \in \mathcal{T}.
\end{aligned}
\end{equation}

Additionally, motivated by~\cite{schaul2015prioritized}, we utilize a prioritized experience replay scheme, in which the absolute TD-error $|\delta_{m,k}|$ was used for evaluating the probability of the $k$-th sampled transition in the mini-batch. Then, the probability of sampling the $k$-th transition is expressed as
\begin{equation}
\begin{aligned}
P_{m,k} = \frac{(|\delta_{m,k}|+\varepsilon)^{\beta}}{\sum_{k'=1}^{K}(|\delta_{m,k}|+\varepsilon)^{\beta}},~\forall m \in \mathcal{M},
\end{aligned}
\end{equation}
where $K$ is the size of mini-batch, $\varepsilon$ is a positive constant value, and $\beta$ is 0.6. Thus, the loss function \textcircled{\small{10}} of the agent $m$ is defined as
\begin{equation}\label{loss}
\begin{aligned}
L(\theta^{Q^m}) = \mathbb{E}\big[\frac{1}{(K \cdot P_{m,k})^{\mu}}(\delta_m)^2\big],
\end{aligned}
\end{equation}
where $\mu$ is given as 0.4.

Then, the critic network \textcircled{\small{8}} of agent $m$ can be updated by the loss function \textcircled{\small{10}} provided in (\ref{loss}). Furthermore, the actor network \textcircled{\small{6}} of agent $m$ can be trained by the policy gradient \textcircled{\small{11}} defined as
\begin{equation}\label{policy}
\begin{aligned}
\nabla_{\theta^{\pi^m}}J = & \mathbb{E}\big[\nabla_{\theta^{\pi^m}}\pi^m(o_{m,t}|\theta^{\pi^m})\\& \nabla_{a_{m,t}}Q^m(s_t,a_t)|\theta^{Q^m}\big],~\forall m \in \mathcal{M}, t \in \mathcal{T}.
\end{aligned}
\end{equation}

Given the UAVs' trajectories, we introduce a low-complexity approach for optimizing the offloading decisions of UEs. Here, we 
do not consider the constraint of the maximal available computing resource in each UAV. This can be readily extended to more practical scenarios, where each UAV can only have a certain amount of the computing resources, with the introduction of the matching algorithm. We will leave this idea for our future work. For each UE in TS $t$, we select the offloading decision based on the following expression
\begin{equation}
\begin{aligned}[l]\label{offp}
z_{n,m,t} = 
\begin{cases}
1, & m=\underset{m'\in\mathcal{M}'}{\text{argmin}}\{E_{n,m',t}\}, \\
0, & \text{otherwise}. \\
\end{cases}
\end{aligned}
\end{equation}
Specifically, after the movement of UAVs, each UE can select the most suitable UAV for offloading, which consumes the least energy. Otherwise, the UE may execute the task itself. If UE $n$ decides to offload a task to UAV $m$, the computational capacity allocated to UE from the UAV is expressed as
\begin{equation}
\begin{aligned}
f_{n,m,t}=\frac{F_{n,t}}{T^{max}-T^{Tr}_{n,m,t}}.
\end{aligned}
\end{equation}

\begin{algorithm}[htpb!]
	\scriptsize
	\caption{MAT}\label{A}
	\begin{algorithmic}[1]
		\FOR{UAV $m$ in $\mathcal{M}$}
		\STATE Initialize actor network $\pi^m(\cdot)$, critic network $Q^m(\cdot)$ with parameters $\theta^{\pi^m}$ and $\theta^{Q^m}$; \
		\STATE Initialize target networks $\pi^{m'}(\cdot)$ and $Q^{m'}(\cdot)$ with parameters $\theta^{\pi^{m'}}=\theta^{\pi^m}$ and $\theta^{Q^{m'}}=\theta^{Q^m}$; \
		\STATE Initialize experience replay buffer $B_m$; \
		\ENDFOR
		
		\FOR{Episode = 1,2,...,$e^{max}$}
		\FOR{UAV $m$ in $\mathcal{M}$}
		\STATE Initialize observation $o_{m,t}$;\
		\ENDFOR
		
		\FOR{TS $t$ in $\mathcal{T}$}
		\STATE Obtain $s_t$;\
		\FOR{UAV $m$ in $\mathcal{M}$}
		\STATE Obtain action $a_{m,t} = \pi^m(o_{m,t}|\theta^{\pi^m})+\epsilon$; \
		\STATE Execute $a_{m,t}$. Note that the UAV will stay at the current location if it flies out of the target area or it is collided with another UAV;\
		\ENDFOR
		\STATE Obtain $a_t$;\
		\FOR{UE $n$ in $\mathcal{N}$}
		\STATE Obtain the available offloading decision $z_{n,m,t}$ that consumes the least energy according to~(\ref{offp});\
		\STATE Calculate $E_{n,m,t}$;\
		\ENDFOR
		\FOR{UAV $m$ in $\mathcal{M}$}
		\STATE Obtain $r_{m,t}$ according to (\ref{rew_fun}) ;\
		\STATE Obtain $o_{m,t+1}$;\
		\ENDFOR
		\STATE Obtain $s_{t+1}$;\
		\FOR{UAV $m$ in $\mathcal{M}$}
		\STATE Store transition $\{s_t,a_t,r_{m,t},s_{t+1}\}$ into experience replay buffer $B_m$ with priority $|\delta_{m}|+\varepsilon$;\
		\IF{learning process starts}
		\STATE Sample a mini-batch of $K$ transitions from $B_{m}$ with probability $P_{m,k}$;\
		\STATE Update critic network according to (\ref{loss});\
		\STATE Update actor network according to (\ref{policy});\
		\STATE Update target networks with updating rate $\tau$:

		\quad $\theta^{\pi^{m'}} \leftarrow \tau\theta^{\pi^m}+(1-\tau)\theta^{\pi^{m'}}$; \

		\quad $\theta^{Q^{m'}} \leftarrow \tau\theta^{Q^m}+(1-\tau)\theta^{Q^{m'}}$;\
		\STATE Update priorities of $K$ transitions;\
		\ENDIF
		\ENDFOR
		\ENDFOR
		\ENDFOR
	\end{algorithmic} 
\end{algorithm}

We provide the pseudo code of proposed procedure in Algorithm~\ref{A}. Specifically, we carry out the initialization between Line 1 and 5 at the beginning, where each UAV initializes its actor, critic and two target networks. Then, the training procedure starts from Line 6, where each UAV first obtains its observation from the environment \textcircled{\small{1}}. Note that each UAV is controlled by its dedicated agent \textcircled{\small{2}}. Then, based on the achieved observation, each UAV selects the action $a_{m,t}$, which is generated by its actor network \textcircled{\small{6}}. In order to achieve a better exploration, we add a noise parameter $\epsilon$, which follows a normal distribution with zero mean and a variance of 1. The exploration noise decays with the rate of 0.9995. Then, the UAV executes the action. Note that the UAV will stay at the current location and obtains a penalty $p_m$, if the next location is obtained outside the target area or the UAV is collided with other UAVs. Then, UE selects the UAV which consumes the least energy according to~(\ref{offp}). Next, we obtain the reward $r_{m,t}$ and the next observation $o_{m,t+1}$. Then, each UAV stores the transition \textcircled{\small{3}} into its experience replay buffer $B_m$ \textcircled{\small{4}}. From Line 28 to 34, when the learning procedure starts, the mini-batch \textcircled{\small{5}} with prioritized experience replay \textcircled{\small{12}} scheme samples $K$ transitions from $B_m$. Furthermore, the critic network \textcircled{\small{8}} is updated by the loss function \textcircled{\small{10}} provided in (\ref{loss}), and the actor network \textcircled{\small{6}} is also updated by the policy gradient \textcircled{\small{11}} provided in (\ref{policy}). After that, the pair of target networks are updated at a rate of $\tau$. Finally, we update the priorities of the $K$ sampled transitions.

\section{Simulation Results}
In this section, we rely on our simulations for evaluating the performance of the proposed MAT algorithm. The simulations are conducted by using Python 3.7 and Tensorflow 1.15.0. We employ four fully-connected hidden layers having $[400, 300, 200, 200]$ neurons in both the actor and critic networks. The actor network is trained at the learning rate of $3\times 10^{-5}$, while the critic network is trained at the learning rate of $10^{-4}$. The AdamOptimizer~\cite{kingma2014adam} is used for updating the actor and critic networks. We set the target region to be a square-shaped area with side length of $l^{max} = 100$ m, where 50 UEs are randomly and uniformly distributed. We set the initial coordinates of UAVs to $[10, 10]$, $[90, 90]$, $[10, 90]$ and $[90, 10]$ m. Additionally, each UE generates a single task in each TS. The rest of the parameters can be found in Table~\ref{tab2}.

\begin{table}[htbp!]
	\caption{Simulation parameters}
	\begin{center}
		\begin{tabular}{|l|r|}
			\hline
			\textbf{Notation}&\textbf{Description} \\
			\hline
			$N$& 50\\
			\hline
			$T$& 20\\
			\hline 
			$l^{max}$ & 100 m\\
			\hline
			$D_{n,t}$ & $[10,14]$ Kb\\
			\hline
			$F_{n,t}$ & $[1800,2000]$ cycles/bit\\
			\hline
			$T^{max}$ & 1 s\\
			\hline
			$d^{max}$ & 20 m\\
			\hline
			$R^{max}$ & 20 m\\
			\hline
			$R^{u}$ & 1 m\\
			\hline
			$H$ & 50 m\\
			\hline
			$B$ & 10 MHz\\
			\hline
			$P_n$ & 0.1 Watt\\
			\hline
			$\sigma^2$ & -90 dBm\\
			\hline
			$k_n$ & $10^{-28}$\\
			\hline
			$v_n$ & 3\\
			\hline
			$g_0$ & $1.42\times10^{-4}$\\
			\hline
			$\gamma$ & 0.95\\
			\hline
			$K$ & 256\\
			\hline
			$\tau$ & 0.01\\
			\hline
			$\varepsilon$ & 0.001\\
			\hline
			$B_m$ & $10^5$\\
			\hline
			$e^{max}$ & $3000$\\
			\hline
			$p_m$ & $10$\\
			\hline
		\end{tabular}
		\label{tab2}
	\end{center}
\end{table}

\begin{figure}[htpb!]
	\centering
	\includegraphics[width=3.8in]{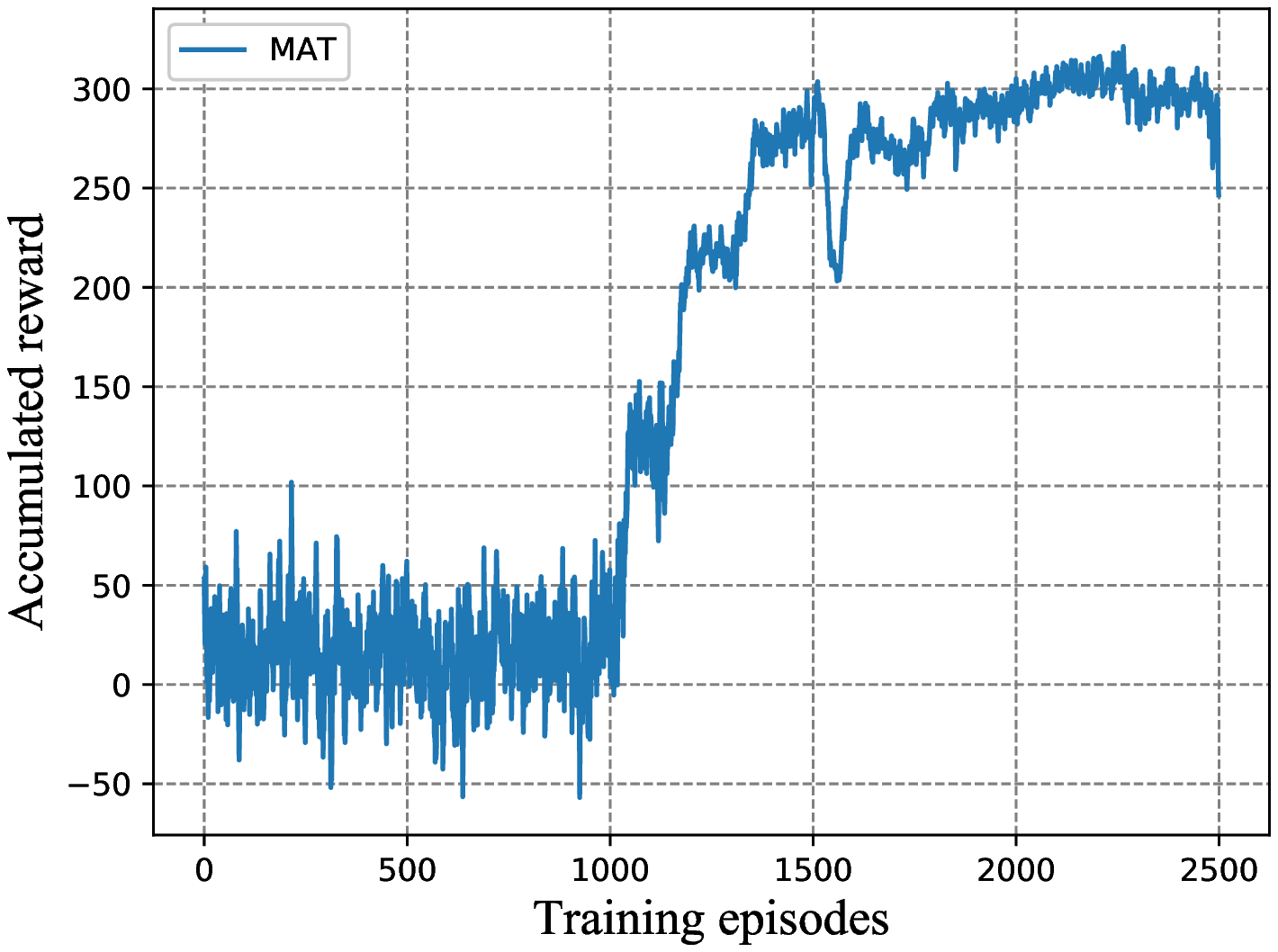}
	\caption{Accumulated reward versus training episodes (with 3 UAVs).	
	}\label{traincurve_u3}
\end{figure}
\begin{figure}[htpb!]
	\centering
	\includegraphics[width=3.8in]{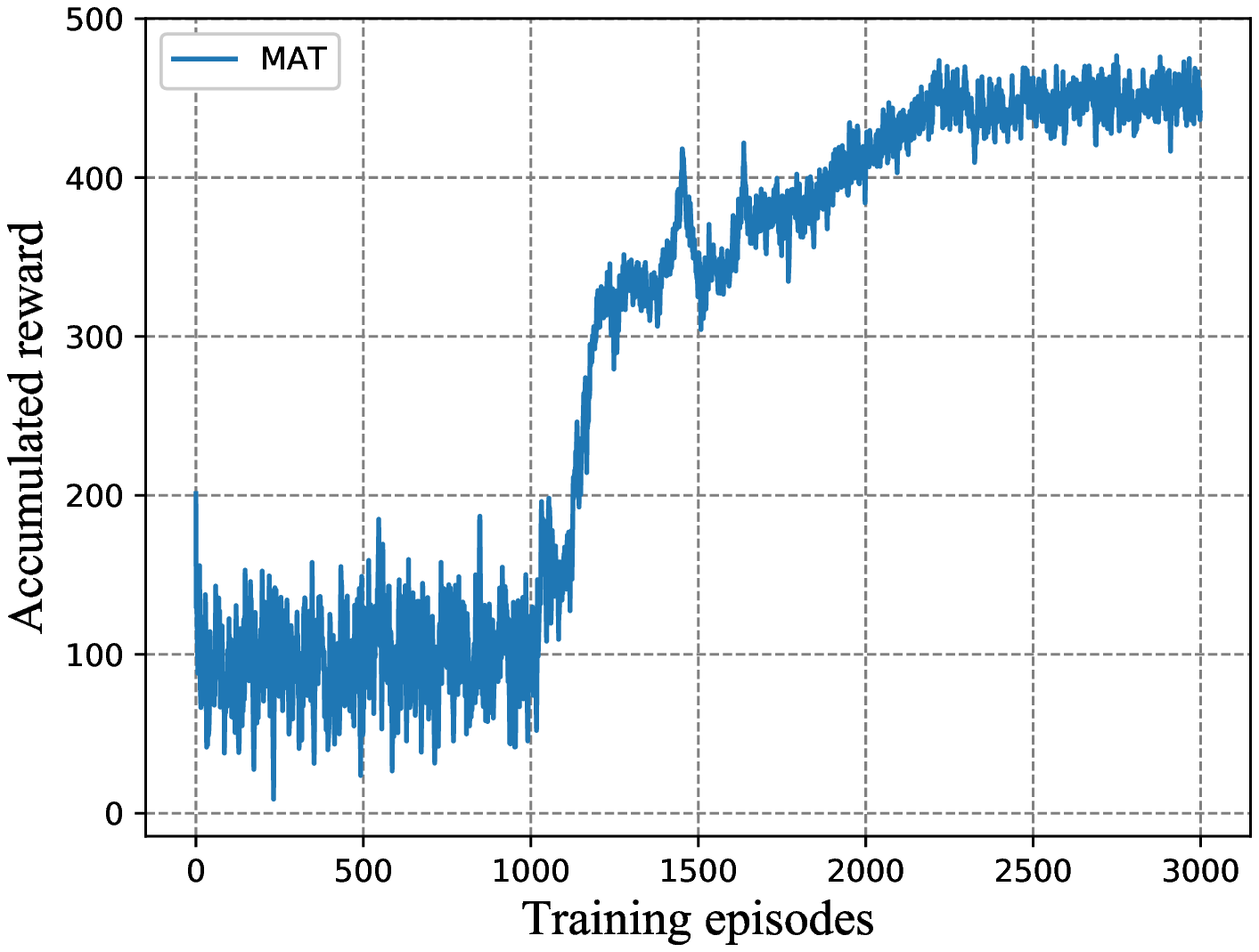}
	\caption{Accumulated reward versus training episodes (with 4 UAVs).	
	}\label{traincurve_u4}
\end{figure}

Firstly, we depict the training curve of MAT in Fig.~\ref{traincurve_u3}, where 3 UAVs are deployed. Observe from Fig.~\ref{traincurve_u3} that the accumulated reward achieved by MAT remains under 50 at the beginning and starts increasing from the 1000-th episode. After about 2000 training episodes, the curve reaches about 300 and then convergence is achieved. 

Then, we increase the number of UAV to 4 and in Fig.~\ref{traincurve_u4}, we depict the accumulated reward achieved by MAT during the training process. Similarly, the curve remains below 200 at the beginning and then increases after the 1000-th episode. It finally saturates around 450. Observe that the accumulated reward seen in Fig.~\ref{traincurve_u4} is higher than that in Fig.~\ref{traincurve_u3}. This is because deploying more UAVs can serve more UEs at the same time, hence resulting in increased accumulated rewards. 

\begin{figure}[htpb]
	\centering
	\includegraphics[width=4in]{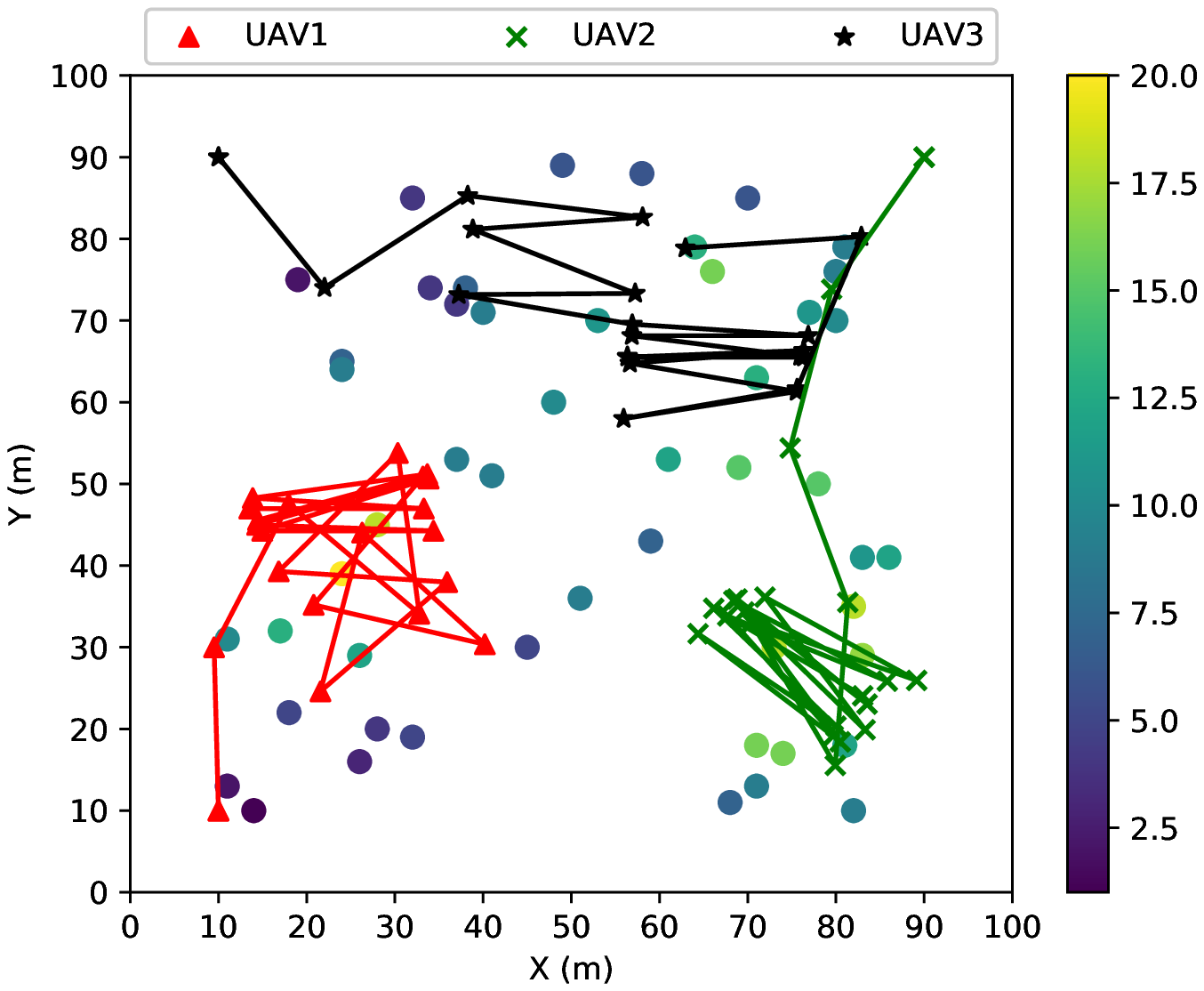}
	\caption{UAVs' trajectories (with 3 UAVs and the locations of UEs are represented by dots.)
	}\label{U3_tra}
\end{figure}

\begin{figure}[htpb]
	\centering
	\includegraphics[width=4in]{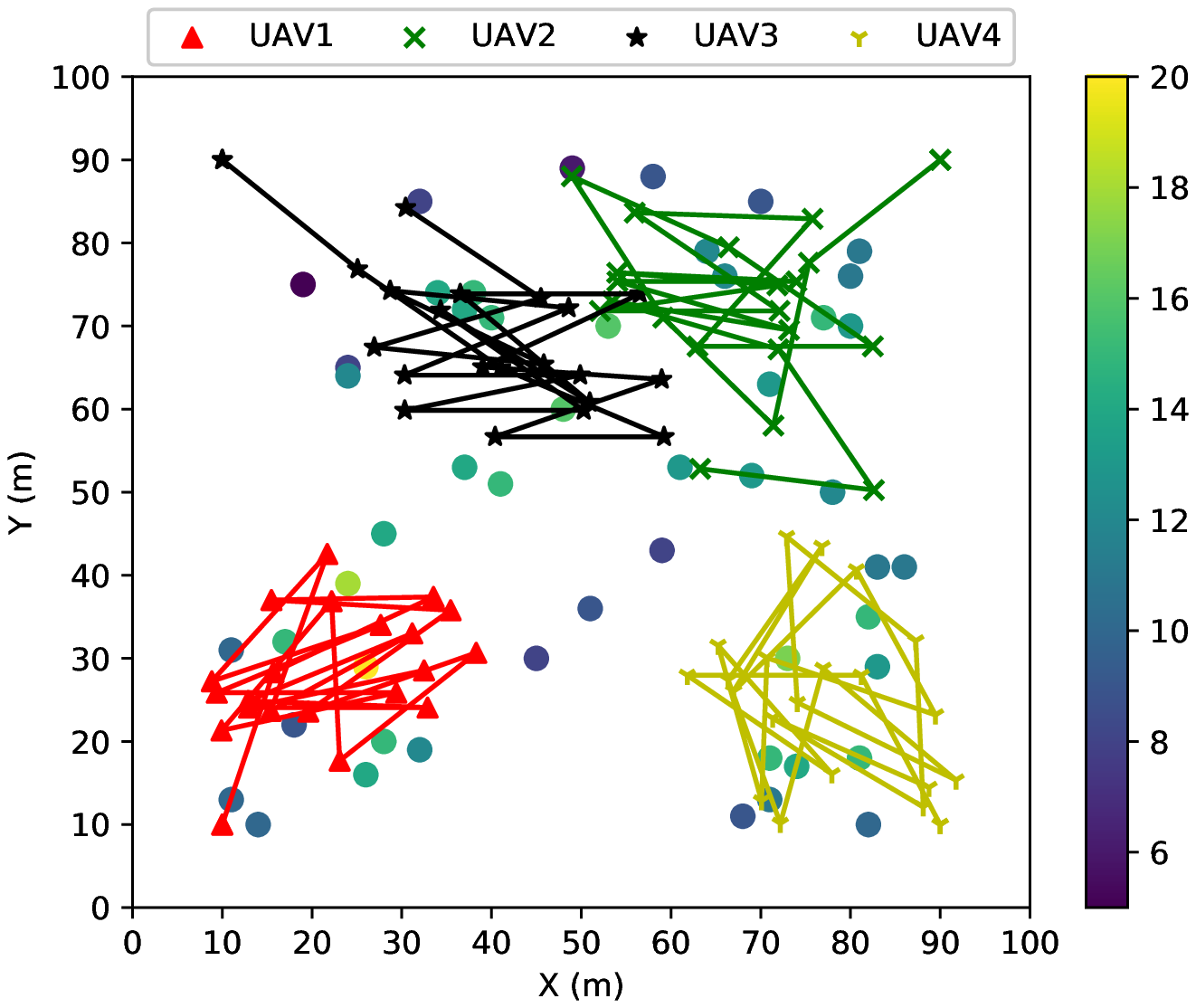}
	\caption{UAVs' trajectories (with 4 UAVs and the locations of UEs are represented by dots.)
	}\label{U4_tra}
\end{figure}

After the training stage, both the model and the network parameters are saved for testing. Next, we compare our algorithm in the cases of 3 and 4 UAVs to the following benchmark solutions:
\begin{itemize}
	\item RANDOM: In this setup, each UAV randomly selects a flying direction within $\alpha_{m,t} \in [0,2\pi)$, and a flying distance $d_{m,t} \in [0,d^{max}]$. Note that the UAVs are restricted to the target area.
	\item CIRCLE: We group all the UEs into a single cluster according to the UEs' coordinates and then all the UAVs fly in a circle twice around the center of the cluster having a radius of $R^{max}$.
\end{itemize}
Note that the MAT, RANDOM, and CIRCLE benchmarks have the same starting points for the UAVs and their offloading decisions are described in Eq. (\ref{offp}).

We first depict the UAV trajectories in Fig.~\ref{U3_tra}, where 3 UAVs are deployed. In this figure, dots represent the location of UEs. We apply a heat map to show the number of times that each UE is served by the UAV commencing from the initial TS to the final TS. The darker the dots, the less amount of time that the UE is spent by the UAV serving. Observe from this figure that all the UAVs move around certain areas, since their coverage range is limited and they have to move for the sake of serving more UEs to increase the fairness index. Additionally, we can see that each UAV covers the particular area in a cooperative manner, so as to maximize the reward defined. For instance, 'UAV2' moves to the lower right corner from its initial location for serving more UEs, while 'UAV3' moves to the upper right corner to help users in this region.

\begin{figure}[htpb]
	\centering
	\begin{subfigure}{.5\textwidth}
		\centering
		\includegraphics[width=1\linewidth]{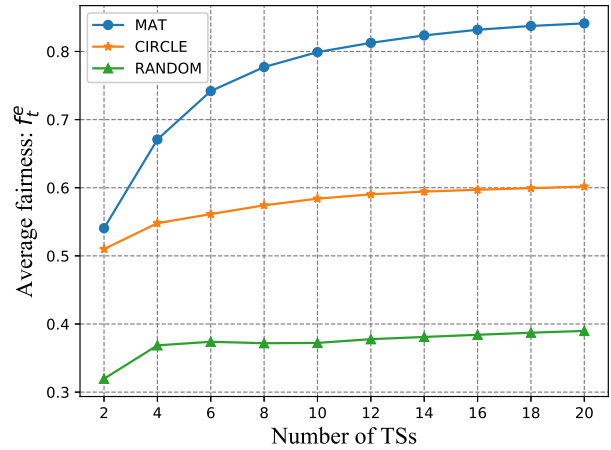}  
		\caption{}
		\label{fe_a_U3}
	\end{subfigure}
	\vskip10pt
	\begin{subfigure}{.5\textwidth}
		\centering
		\includegraphics[width=1\linewidth]{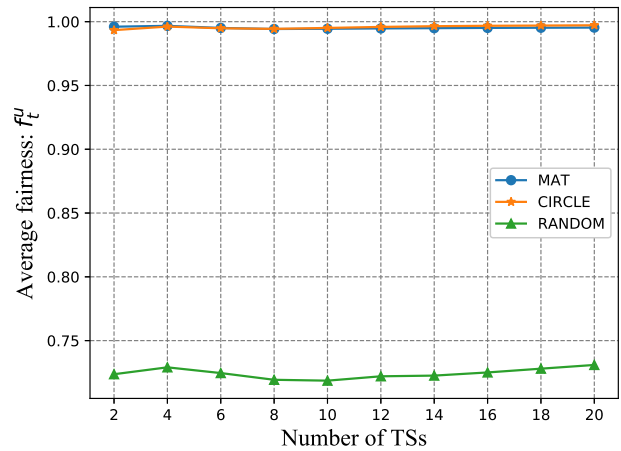}  
		\caption{}
		\label{fu_a_U3}
	\end{subfigure}
	\vskip10pt
	\begin{subfigure}{.5\textwidth}
		\centering
		\includegraphics[width=1\linewidth]{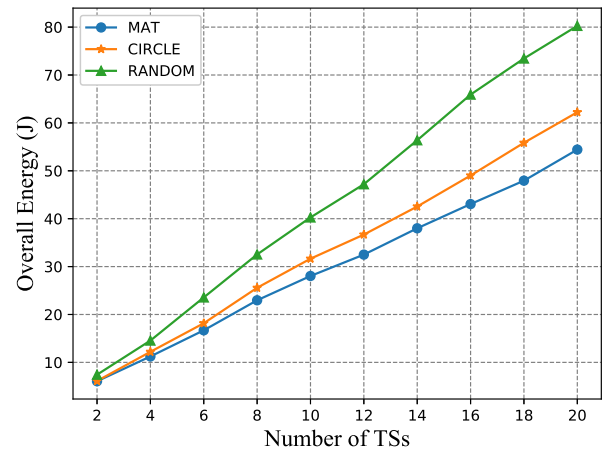}  
		\caption{}
		\label{e_a_U3}
	\end{subfigure}
	\caption{The performance of MAT, CIRCLE and RANDOM versus different number of TSs, in terms of (a) fairness index $f^e_t$, (b) fairness index $f^u_t$ and (c) overall energy consumption of all the UEs (with 3 UAVs).}
	\label{comAllU3}
\end{figure}

\begin{figure}[htpb]
	\centering
	\begin{subfigure}{.5\textwidth}
		\centering
		\includegraphics[width=1\linewidth]{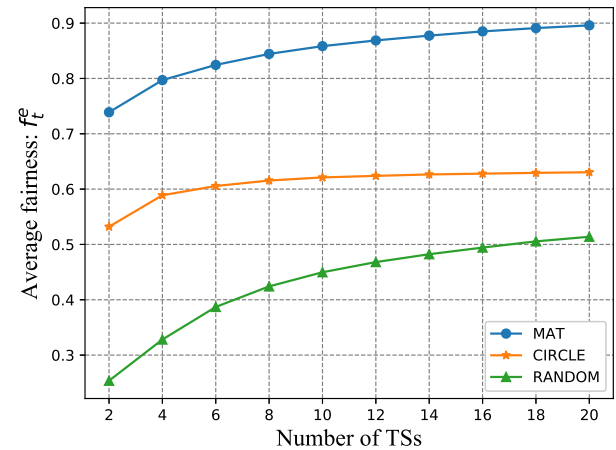}  
		\caption{}
		\label{fe_a_U4}
	\end{subfigure}
	\vskip10pt
	\begin{subfigure}{.5\textwidth}
		\centering
		\includegraphics[width=1\linewidth]{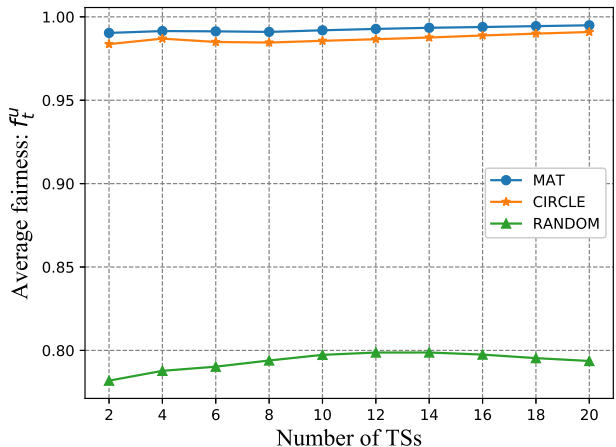}  
		\caption{}
		\label{fu_a_U4}
	\end{subfigure}
	\vskip10pt
	\begin{subfigure}{.5\textwidth}
		\centering
		\includegraphics[width=1\linewidth]{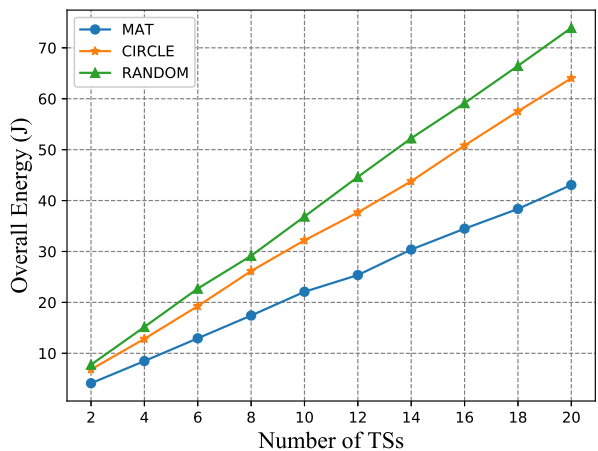}  
		\caption{}
		\label{e_a_U4}
	\end{subfigure}
	\caption{The performance of MAT, CIRCLE and RANDOM versus different number of TSs, in terms of (a) fairness index $f^e_t$, (b) fairness index $f^u_t$ and (c) overall energy consumption of all the UEs (with 4 UAVs).}
	\label{comAllU4}
\end{figure}

Then, we increase the number of UAVs to 4 and depict the trajectories in Fig.~\ref{U4_tra}. Observe that more UAVs result in better coverage. Again, the UAVs cooperate for serving more UEs within the required number of TSs. 
Furthermore, compared to the heat map shown in Fig.~\ref{U3_tra}, 4 UAVs can serve each UE more times than 3. More specially, 4 UAVs can increase the minimum number of serving occurrences from about 2.5 TSs in Fig.~\ref{U3_tra} to about 6 TSs in Fig.~\ref{U4_tra}.

In Fig.~\ref{comAllU3}, we show the fairness attained by 3 UAVs while serving all UEs, the fairness of each UAV's UE-load and the overall energy consumption of all the UEs. Observe from Fig.~\ref{fe_a_U3} that the average fairness $f^e_t$ among all the served UEs achieved by the MAT, CIRCLE and RANDOM regimes increases with the increase of the number of TSs, as expected. Specifically, MAT increases from 0.53 to 0.85, while CIRCLE increases from about 0.5 to 0.6. Finally, RANDOM remains under 0.4.

Then, we show the fairness $f^u_t$ of each UAV's UE-load achieved by the MAT, CIRCLE and RANDOM regimes in Fig.~\ref{fu_a_U3}. Observe that both MAT and CIRCLE approach the fairness of 1, because both solutions can control the UAVs to serve a similar number of UEs. However, RANDOM can only achieve a fairness of 0.75.

Next, in Fig.~\ref{e_a_U3}, we analyse the energy consumed by UEs. We can see that our proposed MAT achieves the best performance, followed by CIRCLE and RANDOM. This is because after training, MAT assists the UAVs in a cooperative way serving the UEs. Hence, more UEs can offload their tasks to UAVs, which results in reduced energy consumption for all the UEs.

Next, in Fig.~\ref{comAllU4}, we increase the number of UAVs to 4 and evaluate the performance of three compared solutions. One can see from Fig.~\ref{fe_a_U4} that the average fairness $f^e_t$  increases with the increase of TSs, as expected. Our proposed MAT can achieve the best performance, reaching at 0.9, whereas the RANDOM performs the worst, which can only achieve about 0.5.

Then, in Fig.~\ref{fu_a_U4}, we draw the fairness of each UAV's UE-load $f^u_t$ achieved by MAT, CIRCLE and RANDOM. One sees that  MAT outperforms CIRCLE and RANDOM, as expected. CIRCLE performs worse than MAT but has much better performance than RANDOM.

Additionally, we show the performance of energy consumed by UEs in Fig.~\ref{e_a_U4}. Similar with before, one can observe that MAT can always achieve the best performance and help UEs to save the energy consumption, while CIRCLE performs the second, followed by RANDOM.
This further proves that with proper training, MAT can control the UAVs to provide better service to UEs.

\section{Conclusions}
In this paper, we have proposed a multi-agent deep reinforcement learning based trajectory control algorithm for jointly maximizing the fairness among all the UEs and the fairness of UE-load of each UAV, as well as minimizing the energy consumption of all the UEs by optimizing each UAV' trajectory and offloading decision from all the UEs. Simulation results show that the proposed MAT has the considerable performance gain over the compared benchmark algorithms.

\bibliographystyle{ieeetran}
\bibliography{reference}

\begin{thebibliography}{10}
\providecommand{\url}[1]{#1}
\csname url@samestyle\endcsname
\providecommand{\newblock}{\relax}
\providecommand{\bibinfo}[2]{#2}
\providecommand{\BIBentrySTDinterwordspacing}{\spaceskip=0pt\relax}
\providecommand{\BIBentryALTinterwordstretchfactor}{4}
\providecommand{\BIBentryALTinterwordspacing}{\spaceskip=\fontdimen2\font plus
\BIBentryALTinterwordstretchfactor\fontdimen3\font minus
  \fontdimen4\font\relax}
\providecommand{\BIBforeignlanguage}[2]{{%
\expandafter\ifx\csname l@#1\endcsname\relax
\typeout{** WARNING: IEEEtran.bst: No hyphenation pattern has been}%
\typeout{** loaded for the language `#1'. Using the pattern for}%
\typeout{** the default language instead.}%
\else
\language=\csname l@#1\endcsname
\fi
#2}}
\providecommand{\BIBdecl}{\relax}
\BIBdecl

\bibitem{zeng2016wireless}
Y.~Zeng, R.~Zhang, and T.~J. Lim, ``Wireless communications with unmanned
  aerial vehicles: Opportunities and challenges,'' \emph{IEEE Communications
  Magazine}, vol.~54, no.~5, pp. 36--42, 2016.

\bibitem{8438896}
Q.~{Wu} and R.~{Zhang}, ``Common throughput maximization in {UAV}-enabled
  {OFDMA} systems with delay consideration,'' \emph{IEEE Transactions on
  Communications}, vol.~66, no.~12, pp. 6614--6627, Dec 2018.

\bibitem{7959158}
L.~{Kong}, L.~{Ye}, F.~{Wu}, M.~{Tao}, G.~{Chen}, and A.~V. {Vasilakos},
  ``Autonomous relay for millimeter-wave wireless communications,'' \emph{IEEE
  Journal on Selected Areas in Communications}, vol.~35, no.~9, pp. 2127--2136,
  Sep. 2017.

\bibitem{fan2018optimal}
R.~Fan, J.~Cui, S.~Jin, K.~Yang, and J.~An, ``Optimal node placement and
  resource allocation for {UAV} relaying network,'' \emph{IEEE Communications
  Letters}, vol.~22, no.~4, pp. 808--811, 2018.

\bibitem{8708930}
H.~{Wang}, J.~{Wang}, G.~{Ding}, J.~{Chen}, F.~{Gao}, and Z.~{Han},
  ``Completion time minimization with path planning for fixed-wing {UAV}
  communications,'' \emph{IEEE Transactions on Wireless Communications},
  vol.~18, no.~7, pp. 3485--3499, 2019.

\bibitem{8685130}
F.~{Cui}, Y.~{Cai}, Z.~{Qin}, M.~{Zhao}, and G.~Y. {Li}, ``Multiple access for
  mobile-{UAV} enabled networks: Joint trajectory design and resource
  allocation,'' \emph{IEEE Transactions on Communications}, vol.~67, no.~7, pp.
  4980--4994, 2019.

\bibitem{lyu2018uav}
J.~Lyu, Y.~Zeng, and R.~Zhang, ``{UAV}-aided offloading for cellular hotspot,''
  \emph{IEEE Transactions on Wireless Communications}, vol.~17, no.~6, pp.
  3988--4001, 2018.

\bibitem{8365881}
J.~{Xu}, Y.~{Zeng}, and R.~{Zhang}, ``{UAV}-enabled wireless power transfer:
  Trajectory design and energy optimization,'' \emph{IEEE Transactions on
  Wireless Communications}, vol.~17, no.~8, pp. 5092--5106, Aug 2018.

\bibitem{9014313}
T.~Q. {Duong}, L.~D. {Nguyen}, H.~D. {Tuan}, and L.~{Hanzo}, ``Learning-aided
  realtime performance optimisation of cognitive {UAV}-assisted disaster
  communication,'' in \emph{2019 IEEE Global Communications Conference
  (GLOBECOM)}, 2019, pp. 1--6.

\bibitem{8727504}
X.~{Liu}, Y.~{Liu}, Y.~{Chen}, and L.~{Hanzo}, ``Trajectory design and power
  control for multi-{UAV} assisted wireless networks: A machine learning
  approach,'' \emph{IEEE Transactions on Vehicular Technology}, vol.~68, no.~8,
  pp. 7957--7969, 2019.

\bibitem{8489991}
J.~{Wang}, C.~{Jiang}, Z.~{Wei}, C.~{Pan}, H.~{Zhang}, and Y.~{Ren}, ``Joint
  {UAV} hovering altitude and power control for space-air-ground {IoT}
  networks,'' \emph{IEEE Internet of Things Journal}, vol.~6, no.~2, pp.
  1741--1753, April 2019.

\bibitem{6863654}
A.~{Al-Hourani}, S.~{Kandeepan}, and S.~{Lardner}, ``Optimal {LAP} altitude for
  maximum coverage,'' \emph{IEEE Wireless Communications Letters}, vol.~3,
  no.~6, pp. 569--572, Dec 2014.

\bibitem{7412759}
M.~{Mozaffari}, W.~{Saad}, M.~{Bennis}, and M.~{Debbah}, ``Unmanned aerial
  vehicle with underlaid device-to-device communications: Performance and
  tradeoffs,'' \emph{IEEE Transactions on Wireless Communications}, vol.~15,
  no.~6, pp. 3949--3963, June 2016.

\bibitem{zeng2016throughput}
Y.~Zeng, R.~Zhang, and T.~J. Lim, ``Throughput maximization for {UAV}-enabled
  mobile relaying systems,'' \emph{IEEE Transactions on Communications},
  vol.~64, no.~12, pp. 4983--4996, 2016.

\bibitem{8247211}
Q.~{Wu}, Y.~{Zeng}, and R.~{Zhang}, ``Joint trajectory and communication design
  for multi-{UAV} enabled wireless networks,'' \emph{IEEE Transactions on
  Wireless Communications}, vol.~17, no.~3, pp. 2109--2121, March 2018.

\bibitem{7918510}
M.~{Alzenad}, A.~{El-Keyi}, F.~{Lagum}, and H.~{Yanikomeroglu}, ``3-{D}
  placement of an unmanned aerial vehicle base station ({UAV}-{BS}) for
  energy-efficient maximal coverage,'' \emph{IEEE Wireless Communications
  Letters}, vol.~6, no.~4, pp. 434--437, Aug 2017.

\bibitem{hu2015mobile}
Y.~C. Hu, M.~Patel, D.~Sabella, N.~Sprecher, and V.~Young, ``Mobile edge
  computing—a key technology towards {5G},'' \emph{ETSI white paper},
  vol.~11, no.~11, pp. 1--16, 2015.

\bibitem{8754787}
K.~{Wang}, P.~{Huang}, K.~{Yang}, C.~{Pan}, and J.~{Wang}, ``Unified offloading
  decision making and resource allocation in {ME-RAN},'' \emph{IEEE
  Transactions on Vehicular Technology}, vol.~68, no.~8, pp. 8159--8172, Aug
  2019.

\bibitem{8016573}
Y.~{Mao}, C.~{You}, J.~{Zhang}, K.~{Huang}, and K.~B. {Letaief}, ``A survey on
  mobile edge computing: The communication perspective,'' \emph{IEEE
  Communications Surveys Tutorials}, vol.~19, no.~4, pp. 2322--2358,
  Fourthquarter 2017.

\bibitem{yang2019energy}
Z.~Yang, C.~Pan, K.~Wang, and M.~Shikh-Bahaei, ``Energy efficient resource
  allocation in {UAV}-enabled mobile edge computing networks,'' \emph{IEEE
  Transactions on Wireless Communications}, vol.~18, no.~9, pp. 4576--4589,
  2019.

\bibitem{8873672}
Y.~{Zhou}, C.~{Pan}, P.~L. {Yeoh}, K.~{Wang}, M.~{Elkashlan}, B.~{Vucetic}, and
  Y.~{Li}, ``Secure communications for {UAV}-enabled mobile edge computing
  systems,'' \emph{IEEE Transactions on Communications}, vol.~68, no.~1, pp.
  376--388, 2020.

\bibitem{motlagh2017uav}
N.~H. Motlagh, M.~Bagaa, and T.~Taleb, ``{UAV}-based {IoT} platform: A crowd
  surveillance use case,'' \emph{IEEE Communications Magazine}, vol.~55, no.~2,
  pp. 128--134, 2017.

\bibitem{7932157}
S.~{Jeong}, O.~{Simeone}, and J.~{Kang}, ``Mobile edge computing via a
  {UAV}-mounted cloudlet: Optimization of bit allocation and path planning,''
  \emph{IEEE Transactions on Vehicular Technology}, vol.~67, no.~3, pp.
  2049--2063, March 2018.

\bibitem{hua2019energy}
M.~Hua, Y.~Wang, Q.~Wu, H.~Dai, Y.~Huang, and L.~Yang, ``Energy-efficient
  cooperative secure transmission in multi-{UAV}-enabled wireless networks,''
  \emph{IEEE Transactions on Vehicular Technology}, vol.~68, no.~8, pp.
  7761--7775, 2019.

\bibitem{8957702}
J.~{Wang}, C.~{Jiang}, H.~{Zhang}, Y.~{Ren}, K.~{Chen}, and L.~{Hanzo},
  ``Thirty years of machine learning: The road to {Pareto}-optimal wireless
  networks,'' \emph{IEEE Communications Surveys Tutorials}, pp. 1--46, 2020.

\bibitem{lecun2015deep}
Y.~LeCun, Y.~Bengio, and G.~Hinton, ``Deep learning,'' \emph{Nature}, vol. 521,
  no. 7553, pp. 436--444, 2015.

\bibitem{sutton1998introduction}
R.~S. Sutton, A.~G. Barto \emph{et~al.}, \emph{Introduction to reinforcement
  learning}.\hskip 1em plus 0.5em minus 0.4em\relax MIT press Cambridge, 1998,
  vol.~2, no.~4.

\bibitem{li2017deep}
Y.~Li, ``Deep reinforcement learning: An overview,'' \emph{arXiv preprint
  arXiv:1701.07274}, 2017.

\bibitem{mnih2015human}
V.~Mnih, K.~Kavukcuoglu, D.~Silver, A.~A. Rusu, J.~Veness, M.~G. Bellemare,
  A.~Graves, M.~Riedmiller, A.~K. Fidjeland, G.~Ostrovski \emph{et~al.},
  ``Human-level control through deep reinforcement learning,'' \emph{Nature},
  vol. 518, no. 7540, p. 529, 2015.

\bibitem{8906180}
J.~{Wang}, C.~{Jiang}, K.~{Zhang}, X.~{Hou}, Y.~{Ren}, and Y.~{Qian},
  ``Distributed {Q}-learning aided heterogeneous network association for
  energy-efficient {IIoT},'' \emph{IEEE Transactions on Industrial
  Informatics}, vol.~16, no.~4, pp. 2756--2764, April 2020.

\bibitem{van2016deep}
H.~Van~Hasselt, A.~Guez, and D.~Silver, ``Deep reinforcement learning with
  double {Q-learning},'' in \emph{Thirtieth AAAI conference on artificial
  intelligence}, 2016.

\bibitem{lillicrap2015continuous}
T.~P. Lillicrap, J.~J. Hunt, A.~Pritzel, N.~Heess, T.~Erez, Y.~Tassa,
  D.~Silver, and D.~Wierstra, ``Continuous control with deep reinforcement
  learning,'' \emph{arXiv preprint arXiv:1509.02971}, 2015.

\bibitem{bu2008comprehensive}
L.~Bu, R.~Babu, B.~De~Schutter \emph{et~al.}, ``A comprehensive survey of
  multiagent reinforcement learning,'' \emph{IEEE Transactions on Systems, Man,
  and Cybernetics, Part C (Applications and Reviews)}, vol.~38, no.~2, pp.
  156--172, 2008.

\bibitem{10.5555/3295222.3295385}
R.~Lowe, Y.~Wu, A.~Tamar, J.~Harb, P.~Abbeel, and I.~Mordatch, ``Multi-agent
  actor-critic for mixed cooperative-competitive environments,'' in
  \emph{Proceedings of the 31st International Conference on Neural Information
  Processing Systems}, ser. NIPS'17.\hskip 1em plus 0.5em minus 0.4em\relax Red
  Hook, NY, USA: Curran Associates Inc., 2017, p. 6382–6393.

\bibitem{7393804}
K.~{Wang}, K.~{Yang}, and C.~S. {Magurawalage}, ``Joint energy minimization and
  resource allocation in {C-RAN} with mobile cloud,'' \emph{IEEE Transactions
  on Cloud Computing}, vol.~6, no.~3, pp. 760--770, July 2018.

\bibitem{yang2013framework}
L.~Yang, J.~Cao, Y.~Yuan, T.~Li, A.~Han, and A.~Chan, ``A framework for
  partitioning and execution of data stream applications in mobile cloud
  computing,'' \emph{ACM SIGMETRICS Performance Evaluation Review}, vol.~40,
  no.~4, pp. 23--32, 2013.

\bibitem{he2017joint}
H.~He, S.~Zhang, Y.~Zeng, and R.~Zhang, ``Joint altitude and beamwidth
  optimization for {UAV}-enabled multiuser communications,'' \emph{IEEE
  Communications Letters}, vol.~22, no.~2, pp. 344--347, 2017.

\bibitem{8274943}
X.~{Lyu}, H.~{Tian}, W.~{Ni}, Y.~{Zhang}, P.~{Zhang}, and R.~P. {Liu},
  ``Efficient admission of delay-sensitive tasks for mobile edge computing,''
  \emph{IEEE Transactions on Communications}, vol.~66, no.~6, pp. 2603--2616,
  June 2018.

\bibitem{bertsekas1995dynamic}
D.~P. Bertsekas, \emph{Dynamic programming and optimal control}.\hskip 1em plus
  0.5em minus 0.4em\relax Athena scientific Belmont, MA, 1995, vol.~1, no.~2.

\bibitem{silver2014deterministic}
D.~Silver, G.~Lever, N.~Heess, T.~Degris, D.~Wierstra, and M.~Riedmiller,
  ``Deterministic policy gradient algorithms,'' in \emph{Proceedings of the
  31st International Conference on International Conference on Machine Learning
  - Volume 32}, ser. ICML’14, 2014.

\bibitem{littman1994markov}
M.~L. Littman, ``Markov games as a framework for multi-agent reinforcement
  learning,'' in \emph{Machine learning proceedings 1994}.\hskip 1em plus 0.5em
  minus 0.4em\relax Elsevier, 1994, pp. 157--163.

\bibitem{schaul2015prioritized}
T.~Schaul, J.~Quan, I.~Antonoglou, and D.~Silver, ``Prioritized experience
  replay,'' \emph{arXiv preprint arXiv:1511.05952}, 2015.

\bibitem{kingma2014adam}
D.~P. Kingma and J.~Ba, ``Adam: A method for stochastic optimization,''
  \emph{arXiv preprint arXiv:1412.6980}, 2014.

\end{thebibliography}

\end{document}